\documentstyle[psfig_sb]{l-aa}              

\def\kev{ke\kern-.15em V}
\def\tkev{\thinspace{ke\kern-.15em V}}
\def\tev{\thinspace{e\kern-.15em V}}

\def\rx{RX J2115.7$-$5840\ }

\def\lap{$\buildrel < \over {_\sim}$} 
\def\gap{$\buildrel > \over {_\sim}$} 

\def\gradi{\ifmmode{^\circ}\else$^\circ$\fi}

\def\kapc{$\kappa_{\rm cyc}$}

\def\rx{RX\,J2115.7--5840}

\def\he2{He\,{\sc ii}\,$\lambda$4686}
\def\h1{He\,{\sc i}\,$\lambda$4471}

\begin{document}
\thesaurus {06(08.02.1, 08.14.2, 08.13.1, 08.09.2 RX\,J2115.7--5840, 02.01.2)}
\title{\rx: a short-period, asynchronous polar\thanks{Based 
in part on observations at the European Southern Observatory
La Silla (Chile) with the 2.2m telescope of the Max-Planck-Society}
}
  \author {
	A.D.~Schwope\inst{1} 
	\and 
	D.A.H.~Buckley\inst{2} 
	\and
        D.~O'Donoghue\inst{3}
	\and
	G.~Hasinger\inst{1}
	\and J.~Tr\"{u}mper\inst{4}
	\and W.~Voges\inst{4}
	}
   \offprints{A.~Schwope {\it (e-mail: ASchwope@aip.de)} }
                               
  \institute {
	Astrophysikalisches Institut Potsdam, An der Sternwarte 16,
	D--14482 Potsdam, FRG
\and South African Astronomical Observatory, PO Box 9, Observatory 7935, Cape 
     Town, RSA
\and  University of Capetown, Dept. Astronomy, Ronde Bosch, 7700, RSA
\and Max-Planck-Institut f\"{u}r Extraterrestrische Physik, D--85740 Garching,
	 FRG
}
\date{Received, accepted}
\maketitle
\markboth{Schwope, A.D. et al.: The new polar \rx }{}
\begin{abstract}
We report phase-resolved optical polarimetric, photometric and spectroscopic
observations of \rx\ (= EUVE J2115--58.6,
Craig 1996) which confirms the system to be a magnetic cataclysmic 
binary of the polar (AM Herculis) subclass.
The optical light curve is sometimes flat and occasionally displays
a pronounced bright phase, reminiscent of the self-eclipse of a small
accretion spot by the revolving white dwarf, as seen in 
self-eclipsing polars. Our period search reveals
ambiguous results only which can be interpret assuming that the white
dwarf is not synchronously rotating with the binary orbit. We find 
circularly polarized cyclotron radiation with $V/I$ ranging from 0\% to 
$-15\%$ on one occasion, from $-8$\% to $+15\%$ on another occasion.
Compared with other polars, the self-occulted accretion 
region of \rx\ had an extreme red 
cyclotron spectrum. In addition, the system has an extreme hard X-ray colour
during the ROSAT all-sky survey observation.
Both properties suggest a low value of the magnetic field strength, and
our best estimate gives $B = 11\pm2$\,MG. Due to the absence of significant 
M-star features in our low-resolution spectra we estimate the minimum 
distance to \rx\ to be $d > 250$\,pc (for an M$5^+$ secondary star).
\end{abstract}

\section{Introduction}
EUVE\,J2115--58.6 was detected during the EUVE all-sky survey with a countrate
of 0.05 cts s$^{-1}$ (Bowyer et al.~1996)
and tentatively identified as a magnetic cataclysmic 
variable by Craig (1996) who identified strong H-Balmer, He{\sc i}, 
He{\sc ii} and Ca{\sc ii} emission lines. Moderate resolution spectroscopy 
in the blue wavelength regime was done by Vennes et al.~(1996). They observed
pronounced radial velocity variations of H$\beta$ and He{\sc ii} and
determined the orbital period of the system of 110.8\,min with a possible
one-cycle-per-day alias of 102.8\,min.

The source was also detected
with the ROSAT-PSPC during the X-ray all-sky survey performed in 1990/1991
at a countrate of 0.380 s$^{-1}$ (galactic
coordinates $l^{II}=337\degr$, $b^{II}=-41\degr$).
We are currently running a program in order to identify optically bright
ROSAT survey sources extracted from the ROSAT All-Sky Survey Bright Source 
Catalog (1RXS, Voges et al.~1997) at high galactic latitudes ($|b^{II}| 
> 30\degr$, limiting countrate 0.2 s$^{-1}$). 
With no certain optical identification at the time, 
\rx\ entered our target list with high priority.
When its nature as magnetic CV became clear from a low-resolution 
spectrum, phase-resolved data were collected in order to study
its main characteristics.
Later it became clear, that
polarimetric and photometric observations has been 
performed coincidentally from SAAO, too. 
We present here the combined results of our optical observations 
obtained over a 70-day basis from South Africa and Chile.
\begin{figure}
\psfig{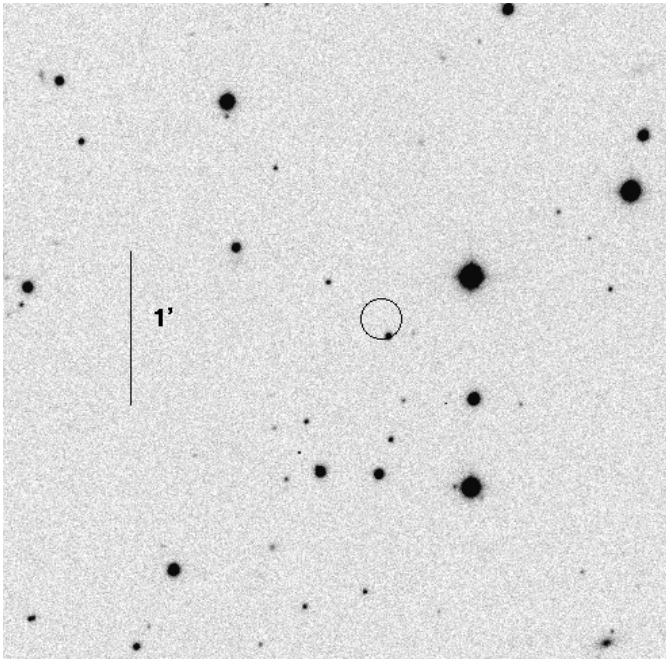}
\caption[chart]{\label{chart}
CCD-image of the field of \rx\ (R-filter), north is top and east to the left.
The circle indicates the X-ray positional error circle, the cataclysmic 
binary is the star on the circle. Its coordinates are $\alpha_{2000} = 
21^h\,15^m 40.6^s, \delta_{2000} = -58\degr 40' 52''$
}
\end{figure}

\section{Observations and analysis}
\subsection{The ROSAT all-sky survey (RASS) observations}
\rx\ was observed during the RASS for a total of 294 sec. 
The source was scanned 13 times with exposures ranging from 16.2 to 25.9 sec.
It showed a modulation of the X-ray flux by 100\% reaching a peak countrate
of 1.2 s$^{-1}$. The mean survey countrate was $0.38\pm0.19$\,s$^{-1}$ and
the mean hardness ratio HR1$= (H -S)/(H+S) = -0.02\pm0.1$, where $H$ and $S$ 
are the counts in the ROSAT hard (0.4 -- 2.4 keV) and soft (0.1 -- 0.4 keV)
bands, respectively. Folded over the most likely optical period, the 
X-ray light curve shows a clear on/off behaviour with length of the X-ray
bright phase extending for $\sim$50\% of the orbital cycle.

With only 102 detected photons the X-ray spectrum of \rx\ is not very well
constrained. It can be fitted using a single bremsstrahlung component
($kT_{\rm br} > 1$\,keV) with X-ray spectral flux at 1 keV of 
2 photons cm$^{-2}$ s$^{-1}$ keV$^{-1}$.  The spectrum is very weakly absorbed,
$N_H = 0.4\pm 1.0 \times 10^{20}$\,cm$^{-2}$, and even compatible with
zero absorption. 

\subsection{Spectroscopic observations}
\begin{figure}[t]
\psfig{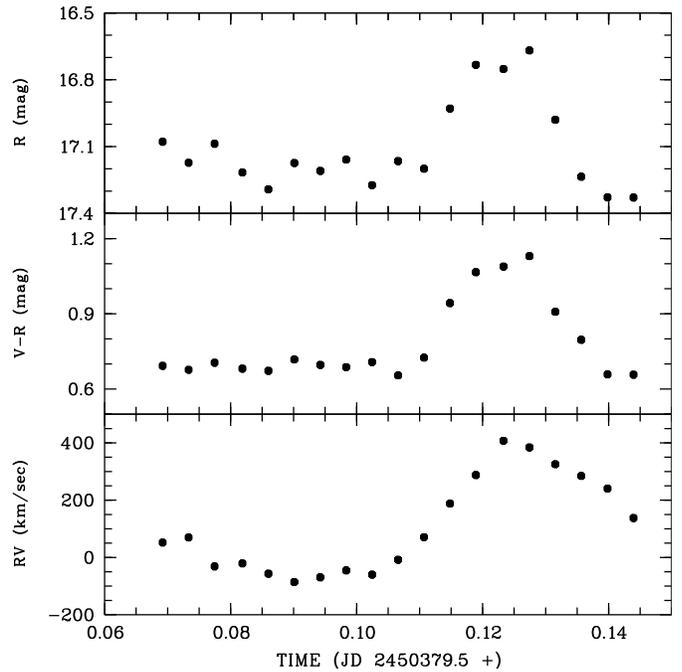}
\caption[chart]{\label{lc_rv}
Optical $R$-band light curve, colour variation $V-R$, and mean radial 
velocity variation derived from low-resolution spectrophotometry obtained
on October 23, 1996
}
\end{figure}
An $R$-band CCD image of the field of \rx\ is reproduced in Fig.~\ref{chart}.
Only one possible optical counterpart of the X-ray source lies within the 
positional error circle of the RASS X-ray observations.
A total of 21 low-resolution spectra of this star (12\,\AA\ 
FWHM, integration time 300 sec)
and one with intermediate resolution (3\,\AA\ FWHM, integration time 600 sec) 
were obtained with the EFOSC2 spectrograph at the ESO/MPG 2.2m-telescope on 
October 23, 1996, between UT 0:19 and 3:30. 

The flux-calibrated low-resolution spectra were folded through Johnson $BVR$ 
filter curves in order to derive broad-band optical light curves. These are 
suspected to be accurate within $\pm$0.5 mag (absolute) but variability and
colours can be determined with much higher accuracy of 
approximately 0.1 -- 0.2 mag. 

The $B$-band light curve does not show any significant variation, the 
$V$-band light curve displays 
marginal variability at a level of 0.2 mag, but the
$R$-band light curve shows a pronounced orbital hump with full amplitude 
of about 0.5 mag centered on HJD 245\,0379.6231. The optical $R$-band light
curve as derived from our low-resolution spectra is shown together with 
the variation in $V-R$ and the radial velocity variation of the main 
emission lines in Fig.~\ref{lc_rv}. Two more spectra (not shown in 
Fig.~\ref{lc_rv}) were recorded during the preceding faint and bright phases, 
respectively. Using all the spectra, a photometric period of $\sim$113 min
was derived by a period search based on Scargles (1982) algorithm.

\begin{figure}[t]
\psfig{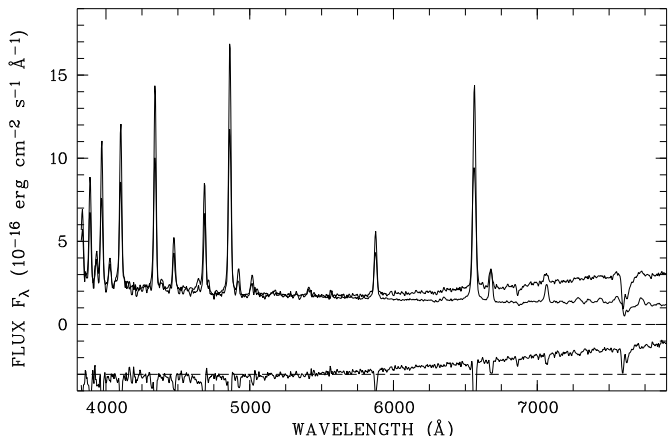}
\caption[chart]{\label{spec_low}
Average bright- and faint-phase low-resolution spectra of \rx\ (12\AA\ FWHM).
The lower curve, which was shifted by three flux units in vertical direction
is the difference of the 
above spectra, regarded as cyclotron spectrum
}
\end{figure}

\begin{figure}[t]
\psfig{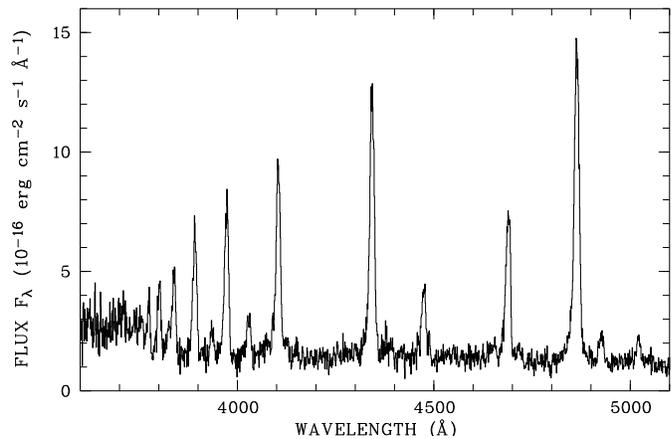}
\caption[chart]{\label{spec_hires}
Single bright-phase spectrum of \rx\ at intermediate spectral resolution 
(3\AA\ FWHM). The spectrum shows the 
typical features of a magnetic cataclysmic 
binary, the Hydrogen Balmer lines up to the series limit as well as lines of
high-ionization species as He{\sc i}, He{\sc ii} and C{\sc iii}/N{\sc iii}
}
\end{figure}

Mean bright- and faint-phase spectra with low spectral resolution 
are shown in Fig.~\ref{spec_low} and
the one spectrum with higher resolution is shown in Fig.~\ref{spec_hires}. 
\rx\  shows the typical features of a magnetic cataclysmic
binary with strong emission lines of the H-Balmer series (including the 
strong Balmer jump in emission), He{\sc i}, He{\sc 
ii}, and the C{\sc iii}/N{\sc iii} Bowen blend at 4640/50\,\AA. The lines 
exhibit a phase-dependent asymmetry and display pronounced radial 
velocity variations. We have determined the radial velocity of 5 main
emission lines (H$\alpha$, H$\beta$, H$\gamma$, He{\sc i}\,5876, He{\sc 
ii}\,4686) by fitting single gaussians. No significant difference between the
radial velocity variations of the different lines was found. We, 
therefore, show the average
radial velocity of the 5 lines in the lower panel of Fig.~\ref{lc_rv}.
Although the emission lines are slightly asymmetric, our resolution is not
high enough in order to discern between possible (likely) multiple 
emission components. Although not reflecting the change of the radial 
velocity variations very well, we used a sine approximation in order to 
estimate the spectroscopic period. The best fit gives $P_{\rm orb} = 114
\pm 4$\,min, in agreement with the period derived from the photometric 
variations of our spectra, and consistent with Vennes et al. (1996) 
spectroscopic period.

\subsection{Photometry}
CCD photometry of \rx\ was undertaken at SAAO over 7 nights, from 1996
September 3 to 9. The observations were made on the 0.75-m telescope with the 
UCT CCD camera, employing a Wright Instruments blue-sensitive EEV CCD chip 
operated in frame transfer mode. Observations were mostly conducted without
a filter, except for those done on 7/8 September, for which alternately a B
and I filter were employed. Details of the observations appear in the
observing log (Table~\ref{obslog}), 
suffice to say that integration times were either
20 or 30 sec for the filterless data, with no dead time between frames.
The observations were obtained in both photometric and non-photometric
conditions, while seeing was equally as variable, from good (\lap 1 arcsec)
to poor (2-3 arcsec). The scale of the CCD is 0.37 arcsec $\rm pixel^{-1}$
in normal mode (which was used for the majority of the observations), and
twice that value for 2 $\times$ 2 prebinning mode, which was used when the 
seeing was poor. The field of the UCT CCD on the 0.75-m telescope is $\rm \sim
2.6 \times 1.8~ arcmin^{2}$.

\begin{figure}[bh]
\psfig{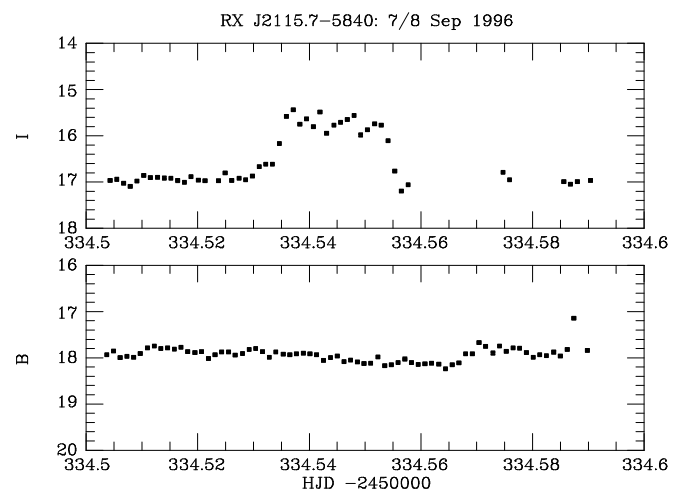}
\caption[]{\label{D1}
CCD-photometry of \rx\ simultaneously obtained in $B$ and $I$-filters
}
\end{figure}

\begin{figure}[ht]
\psfig{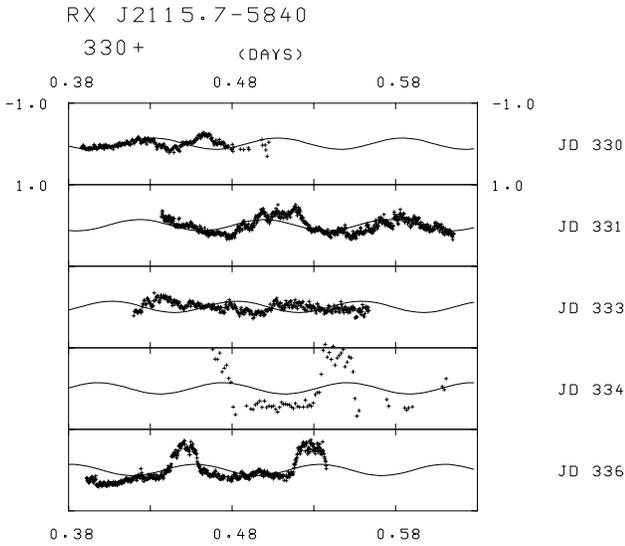}
\caption[]{\label{E2}
CCD-photometry of \rx\ in white light and $I$-filter (JD 334). Time along the
abscissa is given in fractional days, brightness values
were normalized to the mean brightness of each night. Differential white-light
magnitudes with respect to the comparison star directly to the southwest 
of \rx\ for runs 2, 5, 7, and 18
on JD 330, 331, 333, and 336, are $-3.541^m, -3.138^m, -3.210^m,$ and 
$-3.201^m$, respectively. A sine curve connecting all data sets with 
$P = 109.84$\,min, one of our possible photometric periods, is shown
for reference
}
\end{figure}

After the usual flat-fielding and bias subtraction, batch mode DoPHOT
routines (Mateo \& Schechter 1989) were used to obtain both PSF
profile-fitted and aperture magnitudes. Brighter stars on the frames were
used as comparison stars, and differential magnitudes derived. The rms
scatter of the corrected comparison stars was typically 0.006 magnitudes for
the filterless photometry, and $\sim$0.01 mag for the B \& I data. \rx\
exhibits a large degree of variability, up to $\sim 1.4$ mag in I, 1 mag for
filterless (``white-light''), and much less (\lap 0.4 mag) for B. The
pronounced ``hump'' seen in the longer wavelength light curves (I \&
filterless) is not seen in the B light curve (see Fig.~\ref{D1}). Furthermore,
there are substantial night-to-night changes in the light curves, evidenced
by a less pronounced hump, or even a double hump on the first night (see
Fig.~\ref{E2}).

\begin{table*}
\begin{center}
\begin{minipage}{170mm}
\caption{\label{obslog} Observing Log for \rx }
\begin{tabular}{llllccccll}
\hline
\multicolumn{3}{c}{DATE} & \multicolumn{1}{l}{RUN} &
\multicolumn{1}{c}{TELESCOPE} & \multicolumn{1}{c}{HJD START} &
\multicolumn{1}{c}{LENGTH} & \multicolumn{1}{c}{INT} &
\multicolumn{1}{c}{FILTER} & \\
\multicolumn{1}{l}{(day} & \multicolumn{1}{l}{month} &
\multicolumn{1}{l}{year)}	&	&	&(-2450000)	&(h)	&(s) \\
\hline
	&	&   &	&	&		&	& & \\
\multicolumn{3}{l}{\it CCD Photometry:}\\
3/4	&Sep	&96 &W002	&0.75 m	&330.387	&2.78	&30 &w.l. \\
4/5	&Sep 	&96 &W005	&0.75 m	&331.436	&4.32	&20 &w.l. \\
6/7	&Sep 	&96 &W007	&0.75 m	&333.420	&3.46	&30 &B \\
7/8	&Sep 	&96 &W009,W010	&0.75 m	&334.467	&3.46	&60,30 &B,I \\
9/10	&Sep 	&96 &W018	&0.75 m	&336.538	&3.52	&20 &w.l. \\
	&	&   &	&	&		&	& & \\
\multicolumn{3}{l}{\it Polarimetry:} \\
15/16	&Sep 	&96 &P0342	&1.90 m	&342.398	&3.84	&10,180 &w.l. \\
10/11	&Nov 	&96 &P0398	&1.90 m	&398.316	&3.54	&10,180 &w.l. \\
11/12	&Nov 	&96 &P0399	&1.90 m	&399.304	&2.82	&10,180 &w.l. \\
	&	&	&	&	&		&	&  &\\
\hline
\end{tabular}
\end{minipage}
\end{center}
\end{table*}

\subsection{Polarimetry}
White-light photopolarimetry of \rx\ was undertaken on the SAAO 1.9-m
telescope using the UCT Polarimeter (Cropper 1985) on 1997 September 15/16
and November 10/11 and 11/12. The relative faintness of the object precluded
any filtered observations. The instrument was run in the so-called ``Stokes
mode'', simultaneously measuring linear and circular polarization every 180
s, and the intensity every 10 s. Polarization standards were observed at the
end of each night in order to derive the instrumental waveplate offsets,
which are constant to $\pm \sim 2^{\circ}$ from night to night. Sky
measurement were obtained every 15 to 20 min, and the sky background values
interpolated with a polynomial spline before subtraction from stellar
intensity data arrays. The observations were obtained in photometric
conditions with $\sim$ 1 arcsecond seeing, except for the last night (11/12
Nov), for which the seeing was worse (\gap 2 arcseconds).

\begin{figure}[ht]
\psfig{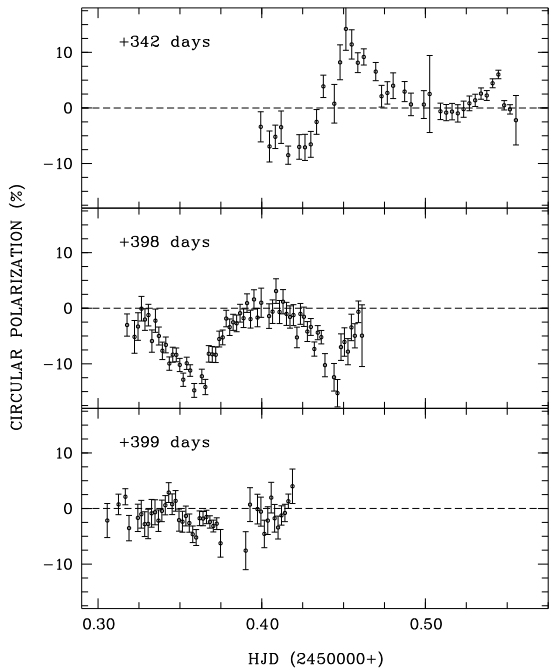}
\caption[]{\label{circpol}
Degree of circular polarization of  \rx\ at the specified days measured 
in white light
}
\end{figure}

\rx\ shows clearly detectable circular polarization, ranging from $\sim -15$ to
$+15\%$ (Fig.~\ref{circpol}). The accuracy of our linear polarization 
measurement was $\sim$3\% and within this accuracy no
clear detection of the linear polarisation is seen.
The various circular polarization curves are quite dissimilar, indicating
that large changes in the accretion geometry between the different observations
took place. The most simple curve at JD 398 is reminiscent of a one-pole 
accreting AM Herculis star with extended self-eclipses of the accretion 
region by the revolving white dwarf. The large positive circular polarization
seen for a rather short phase interval at JD 342 then indicates that also 
a second accretion region on the opposite hemisphere (with opposite
circular polarization polarity) became active.

The light curves recorded simultaneously displays no pronounced orbital hump,
and resemble those seen at JD 330 and 331 (see Fig.~\ref{E2}).

\section{Results and discussion}
The polarimetric observations confirm beyond doubt that \rx\ is indeed a 
magnetic cataclysmic binary, probably of AM Herculis type (polar). 
The borders between polars and intermediate polars (IPs) were blurred 
by the discovery of soft X-ray emitting polarized IPs like 
PQ Gem (other `soft IPs' are discussed by Haberl \& Motch 1995)
on the one hand and asynchronous polars like BY Cam on the 
other hand. However, the degree of synchronism between the white dwarf
and the binary rotation is still an important parameter and we start 
our analysis therefore by a period search using our photometric, polarimetric
and spectroscopic data and compare them  
with the results of Vennes et al.~(1996).

\subsection{Period search}
We subjected the filterless and I-band CCD light curves (Figs.~\ref{D1} and
\ref{E2}) to a period
analysis, using both discrete Fourier transform (DFT) and phase dispersion
minimization (PDM) periodograms. The power spectra show no 
significant features for periods $<$1000 sec, up to the Nyquist frequency. 

\begin{figure}[th]
\psfig{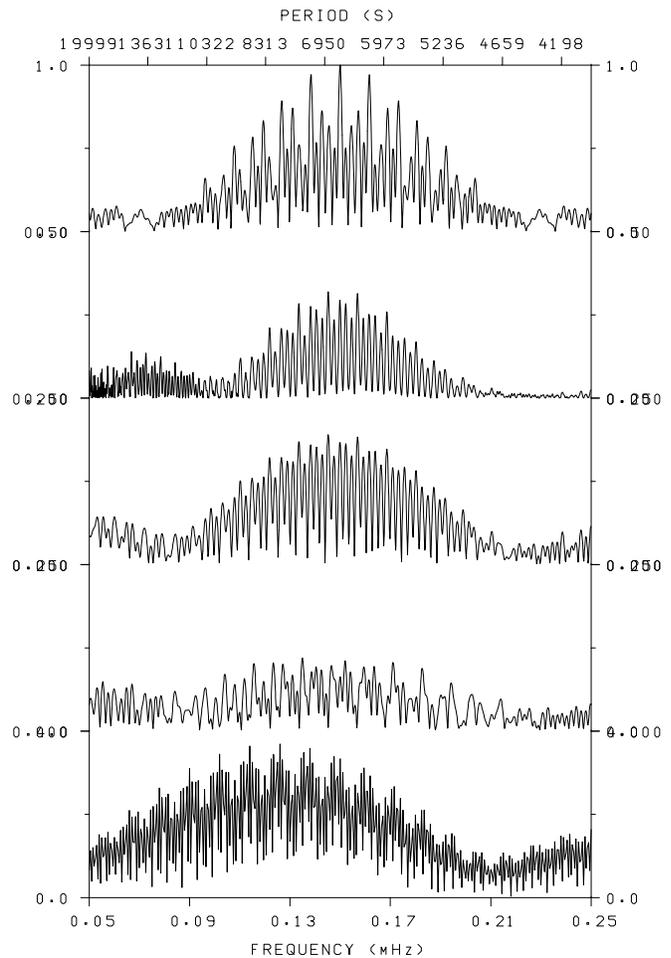}
\caption[]{\label{periodo}
Results of period search of \rx\ using all CCD-data shown in Fig.~\ref{D1}. 
Shown are from top to bottom the window function, the PDM and DFT periodograms,
the DFT periodogram of the photometry after prewhitening by the dominant 
period (114.74 min), and the DFT periodogram of the circular polarization 
data 
}
\end{figure}

Pronounced power appears at $\sim$ 1.5 mHz (110 min), and at the first and
second harmonics, particularly for the DFT, which is more susceptible to
non-sinusoidal waveforms than the PDM periodogram. An inspection of the
periodograms near the dominant frequency indicates that there is not a
``clean'' distribution of frequency peaks with the usual 1 cycle d$^{-1}$
alias structure. This is not surprising given the previously mentioned
nightly changes to the shape of the light curve. In Fig.~\ref{periodo} 
we show the 
periodograms centered near the dominant frequency at
$\sim$0.15 mHz. The strongest peaks in both the DFT and PDM power spectra
occurs at a period corresponding to 114.74\,min (or its alias at 106.27\,min),
although other nearby peaks are only mildly less significant. 
Pre-whitening by the dominant period removes most of the power at 0.15 mHz
(see 4th panel from the top in Fig.~\ref{periodo}), 
indicating that the complex period
structure is likely an artefact of the data sampling. The highest peak for
the prewhitened data occurs at 124\,min, far away from other period 
estimates, the second highest peak at 110\,min is close to the 
spectroscopic period.

After the photopolarimetry runs, we included those
intensity data in our period analysis. Some of these data have poor
quality due to mediocre seeing, and the consequent loss of light from the
small aperture we were forced to use. Inclusion of the polarimetric
intensities did not resolve the ambiguity over the photometric period, and
the periodogram of all the combined photometry (CCD and polarization
intensities) was rather different, with the dominant peak occuring at 98.6
min, and its aliases, none of which coincide with either the spectroscopic
or previously determined photometric periods.
A periodogram of only the $V/I$ circular polarization data (lowest panel
of Fig.~\ref{periodo}) shows highest power at a period of 131.3\,min, which 
is inconsistant with all other period determinations.
However, it may be significant that one of the aliases 
(at 111.8 min) is very close to the purported orbital spectroscopic 
period (110.8 min) reported by Vennes et al. (1996).

The periodograms are clearly affected by the coming and going of the 
bright phase. This becomes evident if one uses for the period search 
only those datasets, where an orbital hump is clearly detectable
(JD 334, 336, and 379, the times of hump center at these three occasions 
can be found in Tab.~\ref{hump_center_times}). 
The corresponding periodogram has power mainly 
at 109.84\,min (566 cycles between days 336 and 379).

If one tentatively assumes that the accretion geometry is the same 
when the circular polarization curve looks most simple 
(JD 398) and when the photometric light curve pattern shows a pronounced hump
(JD 334, 346, and 379) and performs a period search for the times of mid-hump
at these four occasions, a period of 109.65\,min emerges as a 
possible solution (567 cycles between days 336 and 379). 

The different period estimates can be compared to the spectroscopic period, as
derived from the emission line radial velocities by Vennes et al. (1996),
which shows two strongly aliased peaks corresponding to periods of 110.8 and
102.8 min. Neither of our estimates coincides exactly with either 
possible value of the spectroscopic period.

On the basis of the present photometric and polarimetric data, 
it seems that the photometric and spectroscopic periods could be 
discordant, indicating that the
system is asynchronous to a small degree ($\sim$1\%). The changing
shape of the light curve may indicate a pole-swapping accretion mode.
If the far (`southern') pole is accreting, selfeclipses by the revolving
white dwarf occur, leading to pronounced orbital modulation of the
optical light. If the near (`northern') pole accretes, the accretion region
possibly never becomes eclipsed and orbital photometric 
modulations are smoothed, also the changes in polarity of $V/I$ supports
the accretion region moving from one pole to the other.

\begin{table}[t]
\begin{center}
\caption{\label{hump_center_times} 
Times of mid-hump of \rx\ at specified dates }
\begin{tabular}{cc}
\hline
Date & HJD \\
& (-2450000)\\
\hline
7/8 Sept 1996 & 334.5444\\
9/10 Sept 1996 & 336.4513\\
9/10 Sept 1996 & 336.5282\\
23 Oct 1996 & 379.6231\\
\hline
\end{tabular}
\end{center}
\end{table}

The difference between the orbital period $P_{\rm orb}$ and the spin 
period of the white dwarf $P_{\rm wd}$ might be estimated assuming 
that $P_{\rm orb}$ is represented by the spectroscopic period of 110.8\,min
and that $P_{\rm wd}$ is represented by one of the photometric periods derived 
for far-pole accretion, 109.84\,min or 109.65\,min. Hence, the degree of 
asynchronism is of the order of $\sim$1\%. A unique determination of 
this quantity requires photometric monitoring of the system over one
beat cycle (which is $\sim$7 days).

\subsection{The cyclotron spectrum of the far pole}
\begin{figure}[th]
\psfig{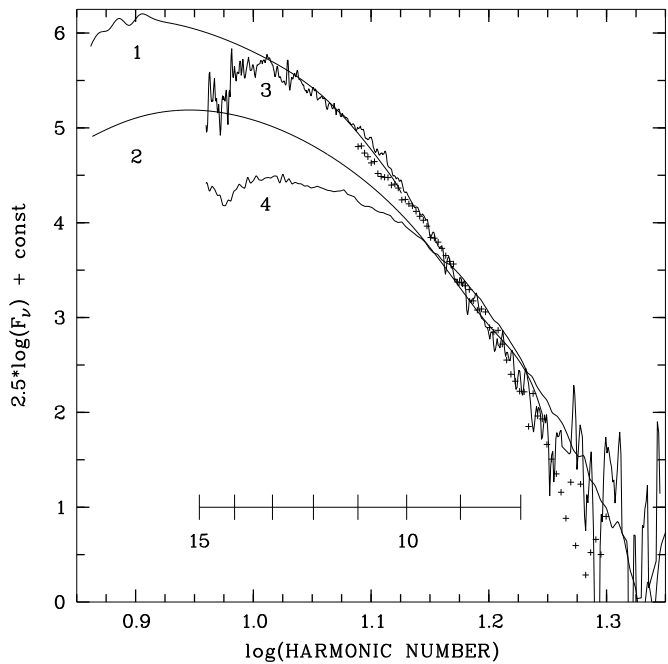}
\caption[chart]{\label{cycspec}
Normalized cyclotron spectra of low-field AM Herculis stars with measured 
field strengths (labelled 1 -- 4) and cyclotron spectrum of \rx\ (represented
by small crosses) for an 
assumed field strength of 11\,MG. Ticks along the horiziontal line indicate, 
where the cyclotron spectrum of \rx\ would have started for field 
strengths $B$ = 15\dots8\,MG. 
The identification of published cyclotron spectra 
of the other AM Hers is: (1) EP Dra -- 16 MG (Schwope \& Mengel 1997),
(2) RXJ1957-57 -- 16 MG (Thomas et al.~1996), (3) and (4) low and
high states of BL Hyi -- 12 MG (Schwope, Beuermann \& Jordan 1995)
}
\end{figure}

As in other selfeclipsing polars (e.g.~Schwope et al.~1995), 
the difference spectrum between the bright and the faint phase can be regarded
as the cyclotron spectrum originating from  the accreting spot active at that
time. This spectrum 
is shown on a linear scale in the lower panel of Fig.~\ref{spec_low} and on 
a logarithmic scale in Fig.~\ref{cycspec}. Also included in Fig.~\ref{cycspec}
are suitably scaled cyclotron spectra 
of other polars which have measured magnetic field strengths. 
The cyclotron spectrum
of \rx\ rises steeply towards long wavelength with peak wavelength 
clearly longward of 8000\,\AA. It does not show any sign of modulation 
by cyclotron harmonics. Since also no Zeeman lines were observed,
no direct measurement of the field strength in \rx\ seems to be possible.
The colour of the cyclotron spectrum indicates that we 
are observing the high-harmonic 
optically thin part of the spectrum which is determined only
by the strong frequency-dependence of the cyclotron absorption coefficient 
\kapc.
Thus at least an estimate for the field strength can be derived,
assuming that the
cyclotron spectra of low-field polars in the optically thin regime 
are similar.


In order to test this hypothesis and use it as tool for the field 
determination of \rx\ we synthesized a common cyclotron
spectrum from low-field polars with measured field strengths 
(Fig.~\ref{cycspec}). 
The absorption coefficient \kapc\ in dimensionless frequency units, 
i.e.~normalized to the cyclotron fundamental, 
is a function of the plasma temperature
and the projection angle only (the latter is the angle between the 
magnetic field and the propagation vectors).
The spread of these parameters must not be too large 
among the different objects for a reliable comparison of their spectra.
The plasma temperatures of the different objects are not yet measured. 
Parameters influencing the plasma temperature are the mass of the white dwarf,
the field strength and the specific mass accretion rate. The former two 
are known to be more or less the same for the objects concerned with here, 
while the latter is widely unknown. For the time being we assume similar values
for the different objects.
We used BL Hyi (Schwope et al.~1995), 
EP Dra (Schwope \& Mengel 1997) and RX\,J1957-57 (Thomas et al.~1996), all of
which have field strengths in the range 12--16 MG. The projection angle at the 
particular phases of the spectra used in the construction of the common
cyclotron spectrum correspond to $\sim$70--80\degr. With the known 
field strength, the observed spectra were  
transformed from wavelength to the dimensionless frequency, normalized
to the cyclotron fundamental frequency. 
The observed spectra peak
at harmonic numbers 8--11, the optically thick Rayleigh-Jeans component
lies in each case in the unobserved infrared spectral regime. 
The different observed turnover frequencies (from being optically thick
to optically thin) reflect different sizes of and densities in the emission 
regions. The different spectra  were then translated to each other
by shifting them vertically until they agree at a certain 
frequency (or harmonic number $m_{H}$).
Constraints for the choice of the finally adopted number, $m_{H} = 14$,
were sufficient low optical thickness (shifts $m_{H}$ to large values)
and sufficient high 
flux in the observed spectra (shifts $m_{H}$ to small values).
The agreement between the different spectra in their usable 
(optically thin) part is surprisingly good, thus justifying our assumptions
on plasma temperature and orientation.

We compared the observed cyclotron spectrum of \rx\ with those of the other 
polars by shifting it along both axes of Fig.~\ref{cycspec} until best 
agreement was reached. A shift along the abscissa corresponds to a change 
of the adopted value of the field strength $B$. A shift along the ordinate
gives just the normalization of the spectrum.
Our best estimate for the field strength 
thus achieved is 11\,MG. We believe that the field strength cannot be much in 
excess of $\sim$13\,MG, because of the observed 
steepness of the cyclotron spectrum of \rx. The field strength 
is probably not much lower than $\sim$9\,MG. The blue end of the cyclotron 
spectrum would then correspond to harmonic numbers as high as and in excess
of $m_{\rm H} \simeq 20$, which has never been observed in any polar 
(due to the negligible power radiated in these high harmonics). We thus 
regard $B =11\pm2$\,MG as a reasonable estimate of the magnetic field strength.

\subsection{A distance estimate}
Our faint-phase low-resolution spectrum does not show any prominent feature
originating from the secondary star. We estimate its contribution to this 
spectrum to be less than 30\% ($\lse 0.55 \times 10^{-16}$ erg cm$^{-2}$
s$^{-1}$ \AA$^{-1}$ at 7540\AA). 
On the assumption that it is a main sequence 
star, which is generally accepted for cataclysmic binaries just below the 
period gap, we may estimate its distance using Bailey's (1981) method
together with the improved calibration by Ramseyer (1994). The major 
uncertainty in this procedure comes from our photometric accuracy and the 
poorly known mass-radius relation of main-sequence stars of late spectral
type. Using e.g.~either the calibration by Caillault \& Patterson (1990) or 
Neece (1984) we obtain the possible spectral type (mass), 
scaled K-brightness (using M-dwarf template spectra), 
surface brighness $S_K$ and stellar radius as 
(Sp, $K$, $S_K$, $\log(R_2/R_\odot)$) = 
(M5$^+$, 15.3, 5.3, -0.61) and 
(M4.5, 16.1, 4.7, -0.73), respectively. With $\log d = ((K-S_K+5)/5 - 
\log(R_2/R_\odot))$ this yields a 
lower limit distance estimate of $d > 250$\,pc for the former, and $d > 
350$\,pc for the 
latter case. Should the spectral type of the secondary be later than 
we have assumned the lower limit distance would be smaller.

\subsection{\rx\ as X-ray and EUV-emitter}
Although \rx\ is a bright, variable source at X-ray wavelengths, it was missed 
by previous identification programmes 
(soft survey: e.g.~Beuermann \& Burwitz 1995, hard survey: Hasinger
et al.~1996) due to its unusual hard X-ray spectrum which places it between
the selection boundaries of these previous identification 
programmes. One may thus speculate about some more low-field polars
as counterparts of relatively hard ROSAT survey sources.

The X-ray spectral shape as seen with ROSAT (although not well determined) 
does not make the assumption of a soft blackbody component necessary, unless
to the case of all other AM Herculis stars. This 
makes the system similar to  the `classical'
intermediate polars (known before ROSAT), which have hard thermal 
bremsstrahlung spectra only. The EUV-detection on the other hand clearly
shows the presence of a soft component. It is not clear, however, if both 
components orginate from the same accretion spot. Perhaps the system has one
IP-like pole emitting predominantly hard X-ray bremsstrahlung which was 
active during the RASS
and a second polar-like accretion region emitting soft and hard X-rays 
which was active during the EUVE sky survey?
With the
present very limited observational data we clearly cannot answer these 
questions and we need more EUV- and X-ray observations with full phase
coverage in both modes of accretion.

\subsection{Conclusions}
We have found clear and unique evidence for the magnetic nature of the 
new cataclysmic variable \rx\ by the detection of strong and variable
circular polarization. In addition, the radial velocity pattern and the 
shape of the optical light curve (when showing the pronounced hump) 
suggest an AM Herculis type nature of this object. We found, however, some
features which do not fit in the most simple picture of a synchronously
rotating polar: (1) there is no consistent photometric and spectroscopic
period; (2) the optical lightcurve is sometimes flat and sometimes strongly 
modulated, which is not related to changes in the mass acrretion rate;
(3) the circular polarization curve is not repeatable, showing either
only one or both signs of polarization.
One possible explanation for these deviations is the presence of a small
asynchronism ($\sim$1\%) between the orbital and the spin periods of the 
white dwarf. The origin of such an asynchronism is unknown. Other polars
with comparable orbital period (hence, mass accretion rate and thus accretion 
torque) and field strength
(hence, synchronisation torque) are not known to show this behaviour, they are
synchronized.
There are three more asynchonous polars known, 
V1500 Cyg (Stockman et al.~1988),
BY Cam (e.g.~Mouchet et al.~1997), and RXJ1940--10 (Patterson et 
al.~1995) with orbital periods of 197, 201, and 202\,min, beat periods
between spin and orbital period of $\sim$14, 7.8, and -49.5 days, respectively.
All these systems are long-period polars, i.e.~they have orbital periods 
above the 2-3 hour period 
gap. For V1500 Cyg the reason for the asynchronism was found in its 1975 
nova explosion. The mechanism behind for the two other systems are unknown.
If confirmed, \rx\ would be the first asynchronous polar below the period gap.
To confirm it's nature is a challenge for observers, to understand it's
physics a challenge for theorists.

\acknowledgements
This work was supported by the BMB+F under grant 50 OR 9403 5.


\begin{thebibliography}{}
\bibitem[]{}
	Bailey J., 1981, MNRAS 197, 31
\bibitem[]{} Beuermann K., Burwitz, V., 1995,
        ASP Conf.~Ser.~85, 99
\bibitem[]{}
	Bowyer S., Lampton M., Lewis J., et al., 
	1996, ApJS 102, 129
\bibitem[]{}
	Caillault J.--P., Patterson J., 1990, AJ 100, 825
\bibitem[]{}
	Craig N., 1996, IAU Circ.~6297
\bibitem[]{} Cropper M., 1985, MNRAS, 212, 709
\bibitem[]{} Haberl F., Motch C., 1995, A\&A 297, L37
\bibitem[]{}
	Hasinger G., Boller T., Fischer J.-U., et al., 
        1996, Astrophysical Letters and Communications, in press
\bibitem[]{} Mateo M., Schechter P., 1989, in Grosbol P.J., Murtagh F. \&
Warmels R.H., eds, 1st ESO/ST-ECF Data Analysis Workshop, p.69
\bibitem[]{}
	Mouchet M., Bonnet-Bidaud J.M., Somov N.N., Somova T.A., 1997, 
	A\&A in press
\bibitem[]{}
	Neece G.D., 1984, ApJ 277, 738
\bibitem[]{}
	Patterson J., Skillman D.R., Thorstensen J., Hellier C., 1995, 
	PASP 107, 307
\bibitem[]{}
	Ramseyer T.F., 1994, ApJ 425, 243
\bibitem[]{}
	Scargle J.D., 1982, ApJ 263, 835
\bibitem[]{}
	Schwope A.D., Mengel S., 1997, Astron.~Nachrichten 318(1), 25
\bibitem[]{}
	Schwope A.D., Beuermann K., Jordan S., 1995, A\&A 301, 447
\bibitem[]{}
	Stockman H.S., Schmidt G.D., Lamb D.Q., 1988, ApJ 332, 282
\bibitem[]{}
	Thomas H.-C., Beuermann K., Schwope A.D., Burwitz V., 1996, 
	A\&A 313, 833
\bibitem[]{} Vennes S., et al., 1996, AJ, 112, 2254
\bibitem[]{}
	Voges W., Aschenbach B., Boller Th., et al., 1997, 
	Astron.~Astrophys., in press
\end{thebibliography}
\end{document}